\documentclass[preprint,12pt]{elsarticle}

\usepackage{amssymb}
\usepackage{amsmath,slashed}
\newcommand{\sumint}[1]{\sum\!\!\!\!\!\!\!\int\limits_{#1}}
\usepackage{combelow}

\usepackage{xcolor}
\usepackage{ulem}
\usepackage{hyperref}
\hypersetup{colorlinks=true, citecolor=blue, urlcolor=blue, citecolor=red}

\begin{document}

\begin{frontmatter}

\title{Signatures of local acceleration of quark-gluon plasma in the dilepton production}
\author[label1]{Aritra Bandyopadhyay}
\author[label1]{Moulindu Kundu}
\author[label1]{Victor E. Ambruș}
\author[label2,label1]{Maxim~N.~Chernodub}
\affiliation[label1]{organization={Department of Physics, West University of Timișoara},
            addressline={Bd.~Vasile Pârvan 4},
            city={Timișoara},
            postcode={300223},
            country={Romania}}

\affiliation[label2]{organization={Institut Denis Poisson, CNRS UMR 7013, Université de Tours, Université d'Orléans},
            addressline={Parc de Grandmont},
            city={Tours},
            postcode={37200},
            country={France}}

\begin{abstract}
Dilepton production is one of the key probes of the Quark–Gluon Plasma (QGP) that encodes the imaginary part of the electromagnetic current–current correlator. We investigate the effect of local acceleration on the dilepton production by treating acceleration as a small perturbation. Using the thermal Dirac propagator in an accelerated frame within the imaginary-time formalism, we compute the photon polarization tensor and extract its imaginary part. Comparison with the zero-acceleration case isolates the distinct contributions of acceleration to dilepton yields.
\end{abstract}


\end{frontmatter}



\section{Introduction}
\label{sec1}
The study of many-particle systems, particularly the strongly interacting QCD medium in heavy-ion collisions (HIC), remains a central theme in high-energy physics. Recent attention has turned to extreme conditions such as high temperature, density, strong fields, isospin asymmetry, rotation, and acceleration. In this work, we focus on acceleration, argued to be significant in the early stages of HIC~\cite{Castorina:2007eb,Prokhorov:2025vak} until the central rapidity plateau forms.

Medium properties are encoded in correlation functions (CFs) and their spectral representations, which determine key observables like the dilepton production rate (DR). In particular, dileptons in the intermediate invariant mass range are of prime interest, as they carry uncontaminated information from the QGP stage. Since lepton pairs escape with minimal final-state interactions, DR serves as a sensitive probe of the early accelerating medium and provides valuable input for space-time evolution models. Importantly, while leptons themselves are unaffected by the acceleration of the medium  due to their large mean free path, the initial quark–antiquark pairs get sensitive to acceleration which puts them into the focus of our analysis.
In what follows, we discuss the formalism (sec.~\ref{sec2}), explore the results (sec.~\ref{sec3}) and describe our conclusions (sec.~\ref{sec4}).

\section{Formalism}
\label{sec2}
The expression for the unpolarized dilepton production rate for two quark flavors, $N_f=2$, is given by~\cite{McLerran:1984ay,Weldon:1990iw}
\begin{align}
\frac{dN}{d^4x d^4q} = \frac{5\alpha_{\rm{em}}^2}{54\pi^2} \frac{n_B(\omega)}{M^2}\left(\frac{1}{\pi}\mathrm{Im}\left[C^\mu{}_{\mu}(q\equiv \{\omega,{\boldsymbol q}\})\right]\right),
\label{dpr_unpolarized_final}
\end{align}
where $\alpha_{\rm em}$ is the electromagnetic coupling constant, $n_B(\omega)$ is the Bose-Einstein distribution function and $M$ is the invariant mass for the produced dilepton.
In this proceeding, we consider the effect of acceleration at the level of the two-point correlation function $C^\mu{}_\mu (q)$~\cite{Bandyopadhyay2025}, which is then expressed in terms of the Dirac propagator in an accelerated frame $S_E^{(\alpha)}$ as follows:
\begin{multline}
    C_{(\alpha)\mu}^{\mu}(q) =  \frac{1}{\beta V}\int d^4x \int d^4x'~ e^{-i q\cdot (x-x')} \\\times~{\sf{Tr}}_{Dfc} \left[\gamma^\mu ~S_E^{(\alpha)}(x,x') ~\gamma_\mu ~S_E^{(\alpha)}(x,x')\right].
\end{multline}
The Euclidean Dirac propagator in a frame accelerated along the $z$-direction is~\cite{Ambrus:2023smm}: 
\begin{align}
S_E^{(\alpha)}(\tau,z;x')=\sum_{j=-\infty}^{\infty} (-1)^j e^{-j\alpha S^{0z}} S_E^{\rm vac}(\tau_{(j)},z_{(j)} ; x'),
\label{eq_S_alpha}
\end{align}
where the spin prefactor enters as $e^{-j\alpha S^{0z}} = \cos \frac{j \alpha}{2} - i \gamma^0 \gamma^3 \sin \frac{j \alpha}{2}$ with $S^{0z} = \frac{i}{2} \gamma^0 \gamma^3$, $\alpha=a/T$ being the ratio of acceleration $a$ and temperature $T \equiv 1/\beta$. In the limit of small acceleration, $\alpha \ll 1$, the modified coordinates in Eq.~\eqref{eq_S_alpha} can be written as follows:
\begin{subequations}
\begin{align}
    \tau_{(j)} &= \tau - j\beta - z j\alpha + O(a^2) \,, \\
z_{(j)} &= z + \tau j\alpha - \frac{\alpha \beta j^2}{2} + O(a^2)\,.
\end{align}
\label{eq_a_small_path}
\end{subequations}
\!\!\! At vanishing acceleration, $\alpha \to 0$, the transformation~\eqref{eq_a_small_path} reduces to the familiar identification, $\tau_{(j)} = \tau - j \beta$ and $z_{(j)} = z$, which implies the compactification of the time variable to a circle of the length $\beta \equiv 1/T$. Finally, the vacuum propagator $S^{\rm vac}_E(x,x')$ of a fermion of mass $m$ in the Euclidean coordinate space is given by:
\begin{align}
	S^{\rm vac}_E(x,x') = \int\frac{d^4 p}{(2\pi)^4} e^{i p \cdot (x-x')} \frac{m-\slashed{p}}{p^2+m^2}\,,
    \label{eq_S_vac}
\end{align}
where $p = ({\boldsymbol{p}},p_4)$ is the 4-momentum in the Euclidean spacetime. Being a function of $(x-x')$, the propagator~\eqref{eq_S_vac} is translationally invariant.

Within an accelerated medium, the propagator depends on the function $(x_{(j)}-x')$, where $x_{(j)} = ({\tau_{(j)}, z_{(j)}, {\boldsymbol{x}}_\perp})$ is the shifted 4-coordinate~\eqref{eq_a_small_path}. Despite the propagator losing the translational invariance, one can perform a Wigner transform in the limit of small acceleration and consequently represent $S^{\rm vac}_E$ in the momentum space by changing variables from $\{x,x'\} \to \{X=\frac{x+x'}{2},\Delta x = x-x'\}$. Given that $\alpha$ is a small parameter ($\alpha \ll 1$), the propagator admits a perturbative expansion in powers of $\alpha$. We thus arrive at the following expression:
 \begin{align}
	S_E^{(\alpha)}(X,\Delta x) = S_E^0(\Delta x) + \alpha S_E^1(X,\Delta x) + \mathcal{O}[\alpha^2].
	\label{eq:Dirac_prop_expanded}
\end{align}
Here $S_E^0$ is the standard finite-$T$ fermion propagator: 
\begin{equation}
    S_E^0 (\Delta x) = \frac{1}{\beta}\sumint{p} 
    e^{ip\cdot \Delta x} \frac{m-\slashed{p}}{p^2 + m^2} \Bigg\vert_{p_\tau = \omega_n}\,,
    \label{eq_SE_alpha0-final}
\end{equation}
with ${\displaystyle \sumint{p}} = \sum_n \int\frac{d^3p}{(2\pi)^3}$ and $\omega_n = (2n+1)\pi/\beta$ being the fermionic Matsubara frequency. The first-order correction $S_E^1$ in the acceleration is: 
\begin{multline}
    S_E^1 (X,\Delta x) = \frac{1}{\beta}\sumint{p} \frac{\partial}{\partial p_\tau}\left[e^{ip\cdot \Delta x}  
    \left\{p_\tau \left(Z+\frac{\Delta z}{2} \right) - p_z \left(\mathcal{T}+\frac{\Delta\tau}{2}\right)+ \frac{\gamma^0\gamma^3}{2}\right\} \right.\\
    \left.\frac{m-\slashed{p}}{p^2 + m^2} \right]_{p_\tau = \omega_n} + \frac{i}{2\beta}\sumint{p}~p_z~ \frac{\partial^2}{\partial p_\tau^2}\left[
    e^{ip\cdot \Delta x} \frac{m-\slashed{p}}{p^2 + m^2} \right]_{p_\tau = \omega_n}.
    \label{eq_SE_alpha11}
\end{multline}

\begin{figure}
    \centering
    \includegraphics[width=0.9\linewidth]{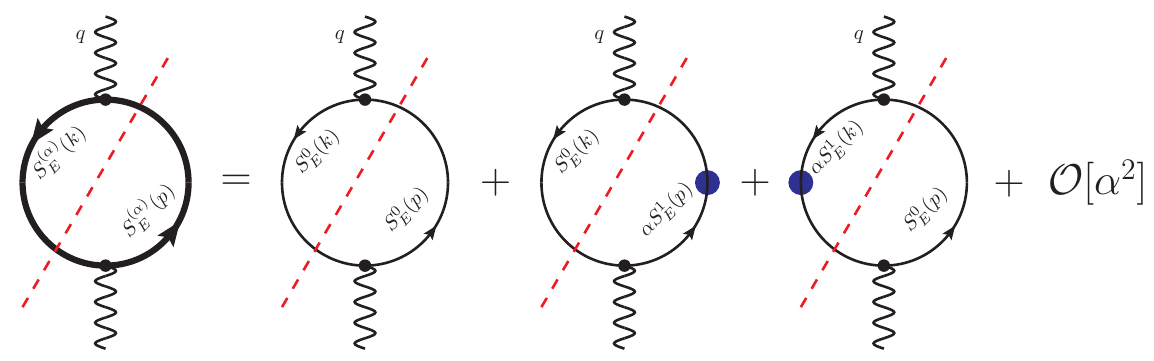}
    \caption{One-loop photon polarization tensor in a hot and weakly accelerated medium expanded in the orders of the acceleration $\alpha$.}
    \label{fig:Cmunu_cut_weak_acc}
\end{figure}

Using Eq.~\eqref{eq:Dirac_prop_expanded} we can subsequently write down the two-point photon correlation function where, by virtue of the Wigner transformation, the Fourier transform is taken only over the relative coordinate $\Delta x$, while the center coordinate $X$ is kept fixed as a slow variable: 
\begin{align}
C^{\mu}_{(\alpha)\mu}(q) &=  \frac{1}{\beta V}\int d^4X \int d^4\Delta x~ e^{-i q\cdot \Delta x}
{\sf{Tr}}_{Dfc} \left[\gamma^\mu S_E^{(\alpha)}(X,\Delta x) \gamma_\mu S_E^{(\alpha)}(X,-\Delta x)\right], \nonumber\\
&= C_{0\mu}^{\mu}(q) + \alpha ~ C_{1\mu}^{\mu}(q) + \mathcal{O}[\alpha^2].
\end{align}
The contribution $C_{0\mu}^{\mu}$ in the absence of acceleration corresponds to the Born rate. On the other hand, $C_{1\mu}^{\mu}$ represents the first-order corrections due to the presence of the weak acceleration in the system (see Fig.~\ref{fig:Cmunu_cut_weak_acc}). 
The evaluation of $C_{0\mu}^{\mu}$ and $C_{1\mu}^{\mu}$ will be presented in detail in~\cite{Bandyopadhyay2025}. In this work, we only give the final results and discuss their consequences in the following section.

\section{Results and Discussion}
\label{sec3}
 
Since the objective of our present study is to observe the effect of acceleration on the dilepton production rate, we show the results for the ratio of the rates in the presence and absence of acceleration. This quantity is equivalent to the ratio of the imaginary part of the electromagnetic current-current correlator:
 \begin{equation}
      \frac{{\rm DR}_{(\alpha)}}{{\rm DR}_{(0)}} \equiv \frac{\mathrm{Im}~C_{(\alpha)\mu}^{\mu}(q)}{\mathrm{Im}~C_{0\mu}^{\mu}(q)} = \frac{\mathrm{Im}~C_{0\mu}^{\mu}(q)+ \alpha \mathrm{Im}~C_{1\mu}^{\mu}(q)}{\mathrm{Im}~C_{0\mu}^{\mu}(q)}.
 \end{equation}  
For $N_c$ colors, the imaginary contributions are respectively given as:
\begin{equation}
    \mathrm{Im}~C^{\mu}_{0\mu}(q) = \frac{1}{2\pi |{\bf q}|\beta} N_c N_f(q^2+2m^2) \log\left[\frac{(e^{-\beta\omega}+e^{-\beta\omega_-})(1+e^{-\beta \omega_+})}{(e^{-\beta\omega}+e^{-\beta\omega_+})(1+e^{-\beta\omega_-})}\right],
\end{equation}
and
\begin{multline}
    \mathrm{Im} C_{1\mu}^{\mu}(q) =\frac{N_c N_f}{2\pi\beta}\Biggl[ \frac{2M^2}{\beta|{\bf q}|^2} \log\left[\frac{(e^{-\beta\omega}+e^{-\beta\omega_-})(1+e^{-\beta \omega_+})}{(e^{-\beta\omega}+e^{-\beta\omega_+})(1+e^{-\beta\omega_-})}\right]\\
    - \frac{\omega^2}{|{\bf q}|} \mathcal{M}(\omega,\omega_+)\left[f_1(\omega_+)+f_1(\omega_-) \right] 
    + \omega\,\mathcal{M}(\omega,\omega_+) \left[f_2(\omega_+) - f_2(\omega_-) \right] \\ 
    - 2(q^2 + 2m^2)\frac{\omega^2}{|{\bf q}|}\frac{\partial}{\partial m^2} \Bigl(\mathcal{M}(\omega,\omega_+) \left(f_2(\omega_+) - f_2(\omega_-) \right)\Bigr) + 2\mathcal{M}(\omega,\omega_+) \\
    \times\Bigl[\omega_+\left(\omega_+-\frac{\omega}{|{\bf q}|}\sqrt{\omega_+^2-m^2}\right)
    +\omega_-\left(\omega_- -\frac{\omega}{|{\bf q}|}\sqrt{\omega_-^2-m^2}\right)\Bigr] \frac{\partial\omega_+}{\partial m^2} \Biggr], 
    \label{eq:ImC1_mumu}
\end{multline}
\begin{figure}[t]
    \centering
    \includegraphics[scale=0.5]{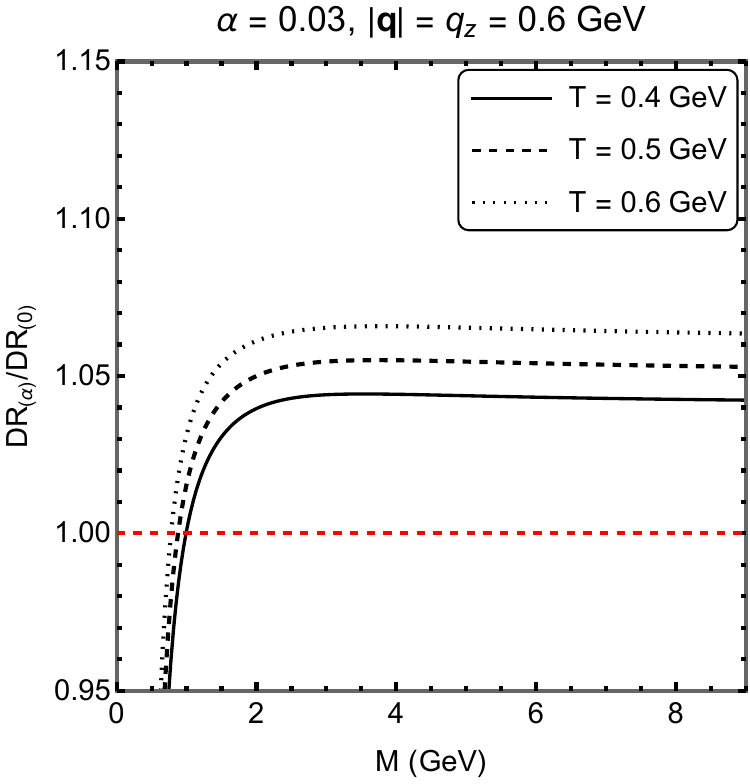} \hspace{0.3cm}\includegraphics[scale=0.5]{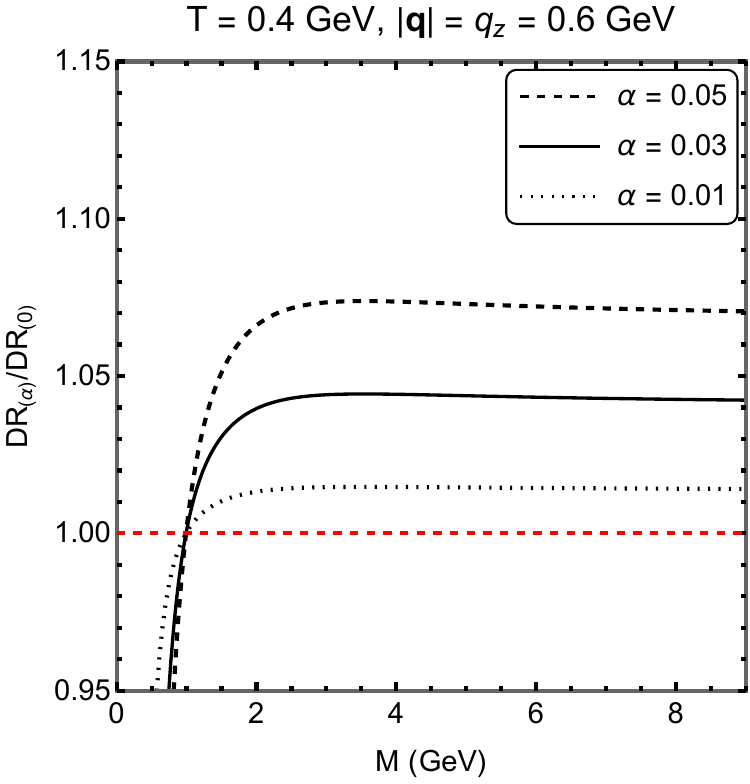}
    \caption{Dilepton production rate as a function of invariant dilepton mass for a weakly accelerating medium (${\rm DR}_{(\alpha)}$) shown in comparison with the Born dilepton rate (${\rm DR}_{(0)}$), (the left panel) at different temperatures $T$ and (the right panel) at various accelerations $a \equiv \alpha T$.}
    \label{fig:DPR-wfa_result}
\end{figure}
where $\mathcal{M}(\omega,x) = 1-n_F(x)-n_F(\omega - x),  f_1(x) = \frac{2(x^2-m^2)^{3/2}}{x q^2 - 2m^2\omega}, f_2(x) = \frac{2(x^2-m^2)x}{x q^2 - 2m^2\omega}$, while $\omega_{\pm} = \frac{1}{2}\left(\omega\pm |{\bf q}|\sqrt{1-\frac{4m^2}{q^2}}\right)$, $n_F(x) = (e^x + 1)^{-1}$ is the Fermi-Dirac distribution function and $q^2 = q_0^2 - \mathbf{q}^2 = M^2$ is the squared invariant mass. At this point, we would also like to mention that the preferred $z$-direction of the accelerated frame breaks the symmetry of the spatial external momentum $\mathbf{q}$ in Eq.~\eqref{eq:ImC1_mumu}. This is not visible explicitly because, for analytical simplicity, we have taken $|\mathbf{q}| = q_z$, i.e. $q_\perp = 0$.

In Fig.~\ref{fig:DPR-wfa_result}, we present the variation of the dilepton production rate (DR) ratio in a weakly accelerated medium as a function of the invariant dilepton mass $M$. The left panel corresponds to a fixed value of $\alpha = 0.03$, where we show the results for three temperatures, $T = 0.4, 0.5, 0.6$ GeV, which in turn correspond to acceleration values of $a = 12, 15, 18$ MeV, respectively. In contrast, the right panel is obtained by fixing the temperature at $T=0.4$ GeV and varying $\alpha = 0.01, 0.03, 0.05$, corresponding to $a = 4, 12, 20$ MeV. The maximum accelerations considered in our analysis are consistent with those theoretically explored in recent Lattice QCD studies~\cite{Chernodub:2024wis}. In both panels, a similar trend is observed: for small invariant masses $M$, DR$_{(\alpha)} < $ DR$_{(0)}$, while beyond a certain cutoff value $M_c$, DR$_{(\alpha)}$ overtakes DR$_{(0)}$ and eventually saturates at a constant ratio DR$_{(\alpha)}/$DR$_{(0)} > 1$, indicating an enhancement of dilepton production in a weakly accelerated medium at higher invariant masses. It is also evident from the results that the cutoff $M_c$ depends on the temperature but is independent of the acceleration $a$. This mass cutoff $M_c$ is dependent on the spatial momentum $|{\bf q}|$ of the dileptons and physically separates two distinct regions. Soft, low-mass dileptons are more sensitive to the expansion, since the acceleration is considered weak; these soft modes are diluted, and their contribution is suppressed, making the weakly accelerated approximation unreliable for very low values of $M$. In contrast, intermediate-mass dileptons, carrying higher momentum, are effectively boosted by the accelerating frame, leading to an enhancement in this region. A similar behaviour in the low-mass region arises in a weakly magnetised medium when the Schwinger proper-time propagator is expanded perturbatively in powers of $eB$~\cite{Bandyopadhyay:2017raf}. This parallel suggests that the issue is generic to weak-field expansions, rather than being specific to acceleration alone.

\section{Conclusion and Outlook}
\label{sec4}

By performing a perturbative expansion with respect to acceleration in a weakly accelerated medium, we demonstrated an enhancement in the dilepton production rate at intermediate invariant masses, which are of particular relevance for the QGP. A more detailed computation, including intermediate steps and extended physical interpretations, will be presented in~\cite{Bandyopadhyay2025}. Possible follow-ups include extending the analysis to arbitrary values of acceleration, incorporating polarization effects of the outgoing lepton pair, and accounting for a spacetime-dependent acceleration profile. Such investigations are currently in preparation. 

\vspace{0.5cm}
\noindent
{\bf Acknowledgements -} This work was funded by the EU’s NextGenerationEU instrument through the National Recovery and Resilience Plan of Romania - Pillar III-C9-I8, managed by the Ministry of Research, Innovation and Digitization, within the project entitled ``Facets of Rotating Quark-Gluon Plasma'' (FORQ), contract no.~760079/23.05.2023 code CF 103/15.11.2022.

\end{document}